\documentstyle[12pt]{article}

\catcode`\@=11
\long\def\@makefntext#1{
\protect\noindent \hbox to 3.2pt {\hskip-.9pt
$^{{\ninerm\@thefnmark}}$\hfil}#1\hfill}		

\def\@makefnmark{\hbox to 0pt{$^{\@thefnmark}$\hss}}  

\def\ps@myheadings{\let\@mkboth\@gobbletwo
\def\@oddhead{\hbox{}
\rightmark\hfil\ninerm\thepage}
\def\@oddfoot{}\def\@evenhead{\ninerm\thepage\hfil
\leftmark\hbox{}}\def\@evenfoot{}
\def\sectionmark##1{}\def\subsectionmark##1{}}

\renewcommand{\thefootnote}{\fnsymbol{footnote}}

\newcounter{sectionc}\newcounter{subsectionc}\newcounter{subsubsectionc}
\renewcommand{\section}[1] {\vspace*{0.6cm}\addtocounter{sectionc}{1}
\setcounter{subsectionc}{0}\setcounter{subsubsectionc}{0}\noindent
	{\normalsize\bf\thesectionc. #1}\par\vspace*{0.4cm}}
\renewcommand{\subsection}[1] {\vspace*{0.6cm}\addtocounter{subsectionc}{1}
	\setcounter{subsubsectionc}{0}\noindent
	{\normalsize\it\thesectionc.\thesubsectionc. #1}\par\vspace*{0.4cm}}
\renewcommand{\subsubsection}[1]
{\vspace*{0.6cm}\addtocounter{subsubsectionc}{1}
	\noindent {\normalsize\rm\thesectionc.\thesubsectionc.\thesubsubsectionc.
	#1}\par\vspace*{0.4cm}}

\newcounter{appendixc}
\newcounter{subappendixc}[appendixc]
\newcounter{subsubappendixc}[subappendixc]

\renewcommand{\appendix}[1] {\vspace*{0.6cm}
        \refstepcounter{appendixc}
        \setcounter{figure}{0}
        \setcounter{table}{0}
        \setcounter{equation}{0}
        \renewcommand{\thefigure}{\Alph{appendixc}.\arabic{figure}}
        \renewcommand{\thetable}{\Alph{appendixc}.\arabic{table}}
        \renewcommand{\theappendixc}{\Alph{appendixc}}
        \renewcommand{\theequation}{\Alph{appendixc}.\arabic{equation}}
        \noindent{\bf Appendix \theappendixc #1}\par\vspace*{0.4cm}}

\def\abstracts#1{{

\centering{\begin{minipage}{12.2truecm}\footnotesize\baselineskip=12pt\noindent
	\centerline{\footnotesize ABSTRACT}\vspace*{0.3cm}
	\parindent=0pt #1
	\end{minipage}}\par}}

\renewenvironment{thebibliography}[1]
	{\begin{list}{\arabic{enumi}.}
	{\usecounter{enumi}\setlength{\parsep}{0pt}
\setlength{\leftmargin 1.25cm}{\rightmargin 0pt}
	 \setlength{\itemsep}{0pt} \settowidth
	{\labelwidth}{#1.}\sloppy}}{\end{list}}

\newcounter{itemlistc}
\newcounter{romanlistc}
\newcounter{alphlistc}
\newcounter{arabiclistc}

\newcommand{\fcaption}[1]{
        \refstepcounter{figure}
        \setbox\@tempboxa = \hbox{\footnotesize Fig.~\thefigure. #1}
        \ifdim \wd\@tempboxa > 6in
           {\begin{center}
        \parbox{6in}{\footnotesize\baselineskip=12pt Fig.~\thefigure. #1}
            \end{center}}
        \else
             {\begin{center}
             {\footnotesize Fig.~\thefigure. #1}
              \end{center}}
        \fi}

\newcommand{\tcaption}[1]{
        \refstepcounter{table}
        \setbox\@tempboxa = \hbox{\footnotesize Table~\thetable. #1}
        \ifdim \wd\@tempboxa > 6in
           {\begin{center}
        \parbox{6in}{\footnotesize\baselineskip=12pt Table~\thetable. #1}
            \end{center}}
        \else
             {\begin{center}
             {\footnotesize Table~\thetable. #1}
              \end{center}}
        \fi}

\def\@citex[#1]#2{\if@filesw\immediate\write\@auxout
	{\string\citation{#2}}\fi
\def\@citea{}\@cite{\@for\@citeb:=#2\do
	{\@citea\def\@citea{,}\@ifundefined
	{b@\@citeb}{{\bf ?}\@warning
	{Citation `\@citeb' on page \thepage \space undefined}}
	{\csname b@\@citeb\endcsname}}}{#1}}

\newif\if@cghi
\def\cite{\@cghitrue\@ifnextchar [{\@tempswatrue
	\@citex}{\@tempswafalse\@citex[]}}
\def\citelow{\@cghifalse\@ifnextchar [{\@tempswatrue
	\@citex}{\@tempswafalse\@citex[]}}
\def\@cite#1#2{{$\null^{#1}$\if@tempswa\typeout
	{IJCGA warning: optional citation argument
	ignored: `#2'} \fi}}

 1
 1
 1

\font\ninerm=cmr9

\begin{document}

\centerline{\normalsize\bf NONVACUUM PSEUDOPARTICLES, QUANTUM TUNNELING}
\baselineskip=16pt
\centerline{\normalsize\bf AND METASTABILITY}
\baselineskip=15pt

\vspace*{0.6cm}
\centerline{\footnotesize JIU-QING LIANG}
\baselineskip=13pt
\centerline{\footnotesize\it Department of Physics, University of
Kaiserslautern, P.O.Box 3049}
\baselineskip=12pt
\centerline{\footnotesize\it 67653 Kaiserslautern, Germany}
\centerline{\footnotesize E-mail: jqliang@gypsy.physik.uni-kl.de}
\vspace*{0.3cm}
\centerline{\footnotesize and}
\vspace*{0.3cm}
\centerline{\footnotesize H.J.W. M\"ULLER-KIRSTEN}
\baselineskip=13pt
\centerline{\footnotesize\it Department of Physics, University of
Kaiserslautern, P.O.Box 3049}
\baselineskip=12pt
\centerline{\footnotesize\it 67653 Kaiserslautern, Germany}
\centerline{\footnotesize E-mail: mueller1@gypsy.physik.uni-kl.de}

\vspace*{0.9cm}
\abstracts{It is shown that nonvacuum pseudoparticles can account for quantum
tunneling and
metastability.  In particular the saddle-point nature of the pseudoparticles is
demonstrated,
and the evaluation of path-integrals in their neighbourhood.  Finally the
relation between
instantons and bounces is used to derive a result conjectured by Bogomolny and
Fateyev.}

\section{Introduction}
\indent
One of the most interesting applications of the Euclidean path-integral
approach is
the study of semi-classical instabilities or tunneling processes as Hawking and
Ross\cite{Hawking} emphasised recently.
Instanton transitions related to the possibility of baryon-- and lepton--number
violation in electroweak theory have attracted widespread attention
\cite{Ringwald}.
It has gradually been realised that vacuum instantons and vacuum
bounces which require vacuum boundary conditions may not be appropriate
for the description of tunneling at finite, nonzero energy \cite{Khlebnikov}.
The investigation of quantum tunneling with a new type of instanton--like
configurations which are characterised by nonzero energy and satisfy
manifestly nonvacuum boundary conditions is therefore of great interest.
In the following we consider the new type of instanton--like and
bounce--like configurations called periodic instantons or periodic
bounces or sphalerons \cite{Khlebnikov,Manton,Liang,Zimmerschied} and
investigate their stability for various
potentials in (1+1) dimensions.  We then calculate their tunneling
effects in (1+0) dimensions for various solvable models since their appropriate
transitions reduce to quantum mechanical tunneling problems
\cite{Liangtwo,Liangthree,Liangfour,Liangfive}.  It is
well-known that the latter also play an important role in the investigation
of the large order behaviour of perturbation expansions. We close therefore
with a discussion
of the Bogomolny--Fateyev relation\cite{Bogomolny} which relates the level
splitting
of one case to the energy discontinuity of a related case, the main idea behind
this being
the fact that the instanton is exactly half of the bounce.
The latter point has been exploited recently in the discussion of duality in
gravity
theory\cite{Hawking}.

\setcounter{footnote}{0}
\renewcommand{\thefootnote}{\alph{footnote}}

\section{Solitons, bounces and   sphalerons on  $S^1$ and their stability}

We recall first the behaviour of vacuum pseudoparticles in (1+0) dimensional
quantum mechanics.  For the double--well potential given by
\begin{equation}
V(\phi) =\frac{\mu^2}{2a^2} (\phi^2-a^2)^2
\label{1}
\end{equation}
the instanton solution is the well-known expression
\begin{equation}
\phi_c = a   \tanh [\mu(\tau + \tau_0)]
\label{2}
\end{equation}
where $\tau = it$ is the Euclidean time. The configuration
$\phi_c$ which corresponds to a transition between the degenerate vacua
has nonzero topological charge and is stable.

The effect of tunneling appears in the level splitting for the potential with
two degenerate minima and in the band structure for the  sine--Gordon
potential with an infinite number of degenerate minima.
In the case of the inverted double--well potential
\begin{equation}
V(\phi) = -\frac{\mu^2}{2a^2}(\phi^2-a^2)^2 + \frac{\mu^2}{2a^2}a^4
\label{3}
\end{equation}
the corresponding classical configuration is the bounce
\begin{equation}
\phi_c = a\sqrt2[{\cosh(\mu\sqrt2\tau)}]^{-1}
\label{4}
\end{equation}
This configuration has zero topological charge and is unstable. The tunneling
effect here
is the decay of the (sometimes called ``false") vacuum state  (metastability) .
                                      Calculating the imaginary part of
  the energy  \cite{Liangtwo} one obtains
\begin{equation}
\Im mE_0 = \frac{4\mu}{g}[\frac{\sqrt2\mu^3}{\pi}]  \exp[-
\frac{4\sqrt2\mu^3}{3g^2}]
\label{4}
\end{equation}
where  $g^2 = \frac{\mu^2}{a^2}$. One should note that the vacuum instanton is
an odd
function of its argument whereas the bounce is even, so that their derivatives
(which classically
represent velocities) are respectively even and odd.  Since these derivatives
are also the
translational zero modes, i.e. wave functions with eigenvalue zero, the
instanton is classically
stable, wheras the bounce is not.

We now consider a field $\Phi(x,t)$ in (1+1) dimensions and static
configuration $\phi(x)$
with Lagrangian density
\begin{equation}
{\cal L} = \frac{1}{2}\partial_\mu\Phi\partial^\mu\Phi  -  U(\Phi)
\label{5}
 \end{equation}
where $U$ is respectively the double--well potential, the inverted double--well
potential or the sine--Gordon potential, i.e.
\begin{eqnarray}
U_1[\Phi] &=& \frac{\mu^2}{2a^2}(\Phi^2 - a^2)^2\nonumber\\
U_2[\Phi] &=& - \frac{\mu^2}{2a^2}(\Phi^2 - a^2)^2 +
\frac{\mu^2}{2a^2}a^4\nonumber\\
U_3[\Phi] &=& 1 + \cos \Phi
\label{6}
\end{eqnarray}
The static solution with finite, nonzero energy is given by
\begin{equation}
\frac{1}{2}(\frac{d\phi}{dx})^2 - U[\phi] = - \frac{1}{ 2}c^2
\end{equation}
where with $ i = 1,2,3$
\begin{eqnarray}
- U_1[0] \leq - \frac{1}{2} {c_1}^2 \leq -U_1[a]  &=& 0 \nonumber\\
-U_2[0]\leq - \frac{1}{ 2} {c_2}^2 \leq - U_2[0]  &=& 0\nonumber\\
-U_3[0] \leq - \frac{1}{2}{c_3}^2 \leq - U_3[\pm \pi] &= & 0
\end{eqnarray}
In these cases
\begin{equation}
\phi= 0 ,\phi = \pm a, \phi = 0,\pm 2\pi ,...
\end{equation}
are trivial (constant) solutions and
\begin{equation}
\phi = \pm a,
\phi = 0,\phi = \pm\pi
\end{equation}
are the corresponding vacuum solutions.   The modulus $k$ of the elliptic
functions with                      $ 0 \leq k \leq 1$  in the three  cases is
respectively given by
\begin{eqnarray}
c_1^2 &=&a^2\mu^2( \frac{1-k^2}{1+k^2})^2\nonumber\\
c_2^2 &=& a^2\mu^2(\frac{1-k^2}{1+k^2})^2\nonumber\\
c_3^2 &= &4 (1-k^2)
\end{eqnarray}
Here  and in the following $k'$ is the complementary elliptic modulus
                                       defined by $k'^2 = 1 - k^2$. The
expressions
$\frac{1}{2}c_i^2 $ can be regarded as energies of the appropriate
pseudoparticles. We now
take  $x \epsilon S^1$, and we demand periodicity of the solutions, i.e. $
\phi(x) = \phi(x+L)$.
The  nontrivial solutions are in the three cases respectively  \cite{Liang}
\begin{eqnarray}
\phi(x) &=& \frac{akb(k)}{\mu} sn[b(k)x,k]  \nonumber\\
\phi(x) &=& s_+(k) dn[\beta(k)x, \gamma] \nonumber\\
\phi(x) &=& 2 \arcsin [k sn(x),k]
\label{7}
\end{eqnarray}
where
\begin{equation}
b(k) = \mu (\frac{2}{1+k^2})^{\frac{1}{2}}, \beta(k) = \frac{\mu}{a}s_+(k),
\end{equation}
and
\begin{equation}
s_+(k) = a \frac{1+k}{\sqrt{1+k^2}}, \gamma^2 = \frac{4k}{(1+k)^2}
\label{8}
\end{equation}
 In the limit $k^2 \rightarrow 1$, i.e. $c^2 \rightarrow 0$, we regain the
vacuum solutions, i.e.
\begin{eqnarray}
\phi (x) = a  \tanh (\mu x) \nonumber\\
\phi (x) = a \sqrt2 [\cosh (\mu\sqrt2 x)]^{-1} \nonumber\\
\phi(x)  = 2 \arcsin[\tanh x]
\end{eqnarray}
whereas in the limit $k^2 \rightarrow 0$  the periodic solutions become the
trivial solutions given
above.
The  periodicity requirement implies certain critical values of L, i.e.
respectively
\begin{equation}
 L =  4n K(k) ,
 L = 2n K(k) ,
 L = 4n K(k)
\end{equation}
where $K(k)$ is the complete elliptic integral of the first kind  defined by
\begin{equation}
K(k) = \int_{0}^{\frac{\pi}{2}}{\frac{d\theta}{\sqrt{1-k^2{\sin{\theta}}^2}}}
\end{equation}
and $n = 1,2,3,...$ (i.e. note that $n = 0$ is excluded). Setting
\begin{equation}
\Phi(x,t) = \phi_c(x) + \sum_m\psi_m(x) e^ {i\omega_mt}
\end{equation}
we  obtain the stability or small fluctuation equation
\begin{equation}
( - \frac{d^2}{dx^2} + U''[\phi_c(x)]) \psi_m(x) = {\omega_m}^2 \psi_m(x)
\end{equation}
with
\begin{equation}
\psi_m(x) = \psi_m(x+L)
\end{equation}
For socalled ``classical stability'' we must have ${\omega}_m^2 \geq 0$.
For the vacuum solutions $\psi_m\propto  \sin \; $ or $ \cos(\frac{2\pi}{L}
mx),m=0,1,2,..., $
these conditions are respectively in the three cases
\begin{eqnarray}
\omega_m^2 &=& 4\mu^2 + \frac{4\pi^2}{L^2}m^2 > 0\nonumber\\
\omega_m^2 &=& 2\mu^2+\frac{4\pi^2}{L^2}m^2 > 0\nonumber\\
\omega_m^2 &=&1 + \frac{4\pi^2}{L^2}m^2 > 0
\end{eqnarray}
It is seen that these conditions are satisfied.
In the case of the trivial solutions
\begin{equation}
\psi_m \propto\sin\; \mbox{or}\; \cos(\frac{2\pi}{L}mx)
\end{equation}
at the critical values of L the stability conditions are respectively
\begin{eqnarray}
\omega_m^2 &=& -2\mu^2 + \frac {4\pi^2}{L^2}m^2
=2\mu^2(\frac{m^2}{n^2}-1)\nonumber\\
\omega_m^2 &=&-4\mu^2 + \frac{4\pi^2}{L^2}m^2
=4\mu^2(\frac{m^2}{n^2}-1)\nonumber\\
\omega_m^2&=& -1 +\frac{4\pi^2}{L^2}m^2 =-1 +\frac{m^2}{n^2}
\end{eqnarray}
where $m=0,1,2,...$ and $n=1,2,3,...$.  We see that for $n>m$: $\omega^2<0$,
i.e. in that case
$\phi_c $ is unstable (a sphaleron).

In the case of the nontrivial solutions the fluctuation equation turns out to
be in each
case a Lam\'e equation\cite{Liang}, i.e.
\begin{equation}
\frac{d^2\psi}{dz^2} +[\lambda - N(N+1)\kappa^2sn^2(z,\kappa)]\psi = 0
\label{9}
\end{equation}
The discrete eigenvalues and eigenfunctions in each of the cases of the
potentials $U_i$, with
$i=1,2,3, $ are:\newline
$U_1$:Here $N=2$, $z=b(k)x $,$\kappa^2 = k^2$ and $\lambda =
\frac{\omega^2+2\mu^2}{b^2(k)}$, and
the solutions are
\begin{eqnarray}
\psi_1&=&sn(z,k)cn(z,k)\nonumber\\
\psi_2&=&sn(z,k)dn(z,k)\nonumber\\
\psi_3&=&cn(z,k)dn(z,k)\nonumber\\
\psi_{4,5}&=&sn^2(z,k)- \frac {[1+k^2\pm\sqrt{1-k^2(1-k^2)}]}{3k^2}
\label{10}
\end{eqnarray}
with respectively the following eigenvalues
\begin{eqnarray}     \omega_1^2&=& \frac{6\mu^2}{(1+k^2)}\nonumber\\
\omega_2^2&=&\frac{6\mu^2k^2}{(1+k^2)}\nonumber\\
\omega_3^2&=&0 \nonumber\\
\omega_{4,5}^2&=&2\mu^2(1\mp\frac{2\sqrt{1-k^2(1-k^2)}}{1+k^2})
\label{11}
\end{eqnarray}
where the second last eigenvalue is seen to be negative and the third is that
of the
zero mode.\newline
$U_2$:Here $N=2$ with $z=\beta(k)x$ and $\kappa^2=\gamma^2=\frac{4k}{(1+k)^2}$
and
 $\lambda= 6+\frac{\omega^2-2\mu^2}{\beta^2(k)}$. In this case the solutions
have
respectively the same form as in the first case but with elliptic modulus
$\gamma$
instead of $k$ and the corresponding eigenvalues are
\begin{eqnarray}
\omega_1^2&=&0\nonumber\\
\omega_2^2&=& -\frac{3\mu^2(1-k)^2}{1+k^2}\nonumber\\
\omega_3^2&=&-\frac{3\mu^2(1+k)^2}{1+k^2}\nonumber\\
\omega_{4,5}^2&=&-2\mu^2\mp2\mu^2\frac{\sqrt{1+14k^2+k^4}}{1+k^2}
\label{12}
\end{eqnarray}
where $\omega_2^2,\omega_3^2, \omega_4^2$are seen to be negative.\newline
$U_3$:Here  $N=1, \kappa = k, z=x, \lambda =\omega^2 +1 $, and the
eigenfunctions are
\begin{equation}
\psi_1=cn(x,k),
\psi_2=dn(x,k),
\psi_3=sn(x,k)
\label{13}
\end{equation}
with respectively the following eigenvalues
\begin{equation}
\omega_1^2 =0,\;
\omega_2^2=k^2-1,\;
\omega_3^2=k^2
\label{14}
\end{equation}
We see that in this case the second eigenvalue is negative.  We also observe
that in each case
the number of negative eigenvalues is odd and some eigenvalues merge with
others
in the limit $k^2 \rightarrow1$.
A negative eigenvalue implies,of course, that the corresponding configuration
is a saddle point.
We mention finally that the supersymmetrised versions of the three models
discussed here
have also been investigated\cite{Kulshreshtha}.

\section{Nonvacuum instantons and tunneling}

We now consider the calculation of the level-splitting for the double-well
potential by summing
contributions originating from nonvacuum instantons and corresponding nonvacuum
instanton--
anti-instanton pairs\cite{Liangthree}.  An analogous calculation can be
performed
for the sine-Gordon potential\cite{Liangfive}.
Finally we consider the limiting cases of high and  low energies, high meaning
here energies approaching
the top of the tunneling barrier.

We consider a scalar field $\phi $ in (1+0)-dimensions with mass = 1 and
Lagrangian
\begin{equation}
{\cal L} = \frac{1}{2}(\frac{d\phi}{dt})^2 - V(\phi)
\end{equation}
with potential
\begin{equation}
V(\phi)=\frac{\mu^2}{2a^2}(\phi^2-a^2)^2
\end{equation}
Integrating the classical equation we obtain with $\tau = it$
\begin{equation}
\frac{1}{2}(\frac{d\phi_c}{d\tau})^2-V(\phi_c)=-E_c
\end{equation}
Integrating we obtain
\begin{equation}
\phi_c=\frac{a kb(k)}{\mu}sn[b(k)(\tau+\tau_0)]
\label{15}
\end{equation}
where we have suppressed the elliptic modulus $k$.
The solution $\phi_c $ has been dubbed  ``periodic instanton''
\cite{Khlebnikov}, ``sphaleron''
 \cite{Manton} and ``bounce'' \cite{Funakubo}.
It is convenient to introduce a new parameter $u$ defined
by
\begin{equation}
k^2=\frac{1-u}{1+u}
\label{8.3}
\end{equation}
with $u= \frac{\sqrt{2E_c}}{a\mu} $ and $
b(k)=\mu(\frac{2}{1+k^2})^{\frac{1}{2}}$.
The Jacobian elliptic function $sn$ has period ${\cal T}=4nK(k)$ for
$n=1,2,3,...$.
Setting $b(k)T=K(k)$ we can define as the analogue of the topological charge
the
quantity
\begin{equation}
Q=\frac{1}{2a}[\phi_c(T)-\phi_c(-T)] =k\sqrt{\frac{2}{1+k^2}}
\end{equation}
We consider the half period part of the solution from $\tau = -T$ to $\tau= +T$
 (and so                       with $\tau_0 = 0$) as the trajectory of the
nonvacuum
instanton (as we prefer to call it).We are interested in the transition
amplitude $A_{+,-}$ for
the transition from one side of the central barrier of the double--well
potential to the other.
We let $|E>_{\pm}$ be  the eigenstates of the same energy $E_0$ in the two
wells (with
minima $\phi_{\pm}$ ) if the presence of the other well is ignored.  The finite
height of
the potential barrier in between splits the $\pm$ degeneracy  so that the
eigenstates
become the odd and even states
\begin{equation}
|E>_{o,e} = \frac{1}{\sqrt2}[|E>_+ \mp  |E>_-]
\end{equation}
with eigenvalues $E_o\pm\frac{1}{2}\triangle E$. The desired amplitude then
becomes
\begin{equation}
A_{+,-}= _+<E|  \exp(-2HT)| \ E>_- = - \exp(-2E_0T)\sinh (T\triangle E)
\label{15.5}
\end{equation}
The problem is to calculate the shift $\bigtriangleup E$.We do this with the
help of the
path-integral method.
In this case the amplitude $A_{+,-}$ can be written
\begin{equation}
A_{+,-}= \int \psi_{E+}(\phi_f)\psi
_{E-}(\phi_i)K(\phi_f,\tau_f;\phi_i,\tau_i)d\phi_fd\phi_i
\label{16}
\end{equation}
where $\tau _f-\tau _i = 2T$.  The kernel K is given by the Feynman
path-integral
\begin{equation}
K(\phi_f, \tau _f;\phi _i,\tau _i)\equiv <\phi_f,\tau_f;\phi_i, \tau_i> = \int
{\cal D}[\phi ]exp [-S]
\label{17}
\end{equation}
where the Euclidean action $S$ is given by
\begin{equation}
S=\int_{\tau _i}^{\tau _f} [\frac{1}{2}(\frac{d\phi}{d\tau})^2+V(\phi)]d\tau
\label{18}
\end{equation}
We write the amplitude as a sum over nonvacuum instanton contributions, i.e.
\begin{equation}
A_{+,-}= \sum_{n=0}^\infty A_{+,-}^{(2n+1)}
\label{19}
\end{equation}
where $ A_{+,-}^{(2n+1)} $ denotes the contribution  of the amplitude for one
nonvacuum instanton and n nonvacuum instanton-anti-instanton pairs.

We consider first the one-nonvacuum--instanton contribution.  We set
\begin{equation}
\phi (\tau)= \phi_c(\tau)+X(\tau)
\label{20}
\end{equation}
where $X(\tau)$ is the deviation of $\phi(\tau)$ from the classical trajectory
$\phi_c$  with
fixed endpoints $X(\tau_i) = X(\tau_f) = 0$. We also write $S=S_c+\delta S$
where
\begin{equation}
S_c = W(\phi(\tau_f), \phi(\tau_i), E) + 2 E_c T
\label{21}
\end{equation}
Here
\begin{equation}
W\longrightarrow \frac{4\mu a^2}{3} (1+u)^{\frac{1}{2}} [E(k) - uK(k)]
\label{22}
\end{equation}
in the limits $\phi(\tau_i)=\phi_i\rightarrow -\tilde{a}$ and
$\phi(\tau_f)=\phi_f\rightarrow \tilde{a}$,
where $-\tilde{a}$ and $\tilde{a}$ are the two middle turning points (where
$\tau_i=-T,\tau_f=T$)
($-\tilde{a}'$ and $\tilde{a}'$ are the two other outside turning points)
and $E(k)$ is the complete elliptic
integral of the second kind.In the one-loop approximation we have
\begin{equation}
\delta S = \int_{\tau_i}^{\tau_f} d\tau [\frac{1}{2} (\frac {dX}{d\tau})^2 +
       X^2(3 \frac{\mu^2}{a^2} {\phi_c}^2 - \mu^2)]
\equiv\int_{\tau_i}^{\tau_f}(X\hat{M}X) d\tau
\label{23}
\end{equation}
where $\hat{M}$ is the small fluctuation operator evaluated at the classical
configuration,i.e.
\begin{equation}
\hat{M} = -\frac{1}{2} \frac{d^2}{d\tau^2} +\mu^2 (\frac{3{\phi_c}^2}{a^2} -1)
\label{24}
\end{equation}
The kernel $K$ defined above is then given by
\begin{equation}
K\equiv exp[-S_c].I
\end{equation}
where
\begin{equation}
I = \int_{X(\tau_i)=0}^{X(\tau_f)=0} {\cal D}\{X\}exp(-\delta S)
\end{equation}
We now have to evaluate the integral $I$. The usual analysis starts as follows.
                                    One expands
the fluctuation $X$ in terms of the complete set of eigenfunctions  $\psi_n$
                                      of the small fluctuation
operator $\hat{M}$  with $\hat{M} \psi_n = \omega_n^2 \psi_n$.  Then
$X=\sum_{n}  C_n\psi_n$ and
\begin{eqnarray}
I&=&\int {\cal D}\{C_n\} \det({\frac {\partial X}{\partial C_n}})
                                                     \exp({-\sum_n {C_n^2
\omega_n^2}})\nonumber\\
&=&\det({\frac{\partial X}{\partial C_n}}) \prod_n
[\frac{\pi}{\omega_n^2}]\nonumber\\
&=&\det({\frac{\partial X}{\partial C_n}})\frac{\pi}{\det {\hat{M}}}\nonumber\\
\label{25}
\end{eqnarray}
Here one expects a problem with the negative eigenvalue of the small
 fluctuation operator obtained above.
However, the boundary conditions $X(\tau_i) = X(\tau_f) = 0$ remove this
negative eigenvalue
$\omega_4^2$. This can be seen as follows. The two boundary conditions imply
the equations
\begin{equation}
-C_2k' + C_4(1-\frac{\triangle_1 +\triangle_2}{3k^2}) +C_5(1-\frac{\triangle_1
-\triangle_2}{3k^2})=0
\end{equation}
and
\begin{equation}
C_2k'+C_4(1-\frac{\triangle_1 +\triangle_2}{3k^2})+C_5(1-\frac{\triangle_1 -
\triangle_2}{3k^2})=0
\end{equation}
where $\triangle_1=1+k^2$ and $\triangle_2 =\sqrt{1-k^2(1-k^2)}$.
 These equations imply $C_2=0$ and
\begin{equation}
C_4(1-\frac{\triangle_1+\triangle_2}{3k^2}) =
-C_5(1-\frac{\triangle_1-\triangle_2}{3k^2})
\end{equation}
{}From the definition of the coefficients $C_n$ we obtain
\begin{equation}
C_4=\int X\psi_4^{\star} d\tau=\int sn^2[b(k)\tau] Xd\tau -\int Xd\tau
                                           \frac{\triangle_1+\triangle_2}{3k^2}
\end{equation}
and
\begin{equation}
C_5=\int X\psi_5^{\star}d\tau=\int sn^2[b(k)\tau] Xd\tau-\int
Xd\tau\frac{\triangle_1-\triangle_2}{3k^2}
\end{equation}
Here $\int_{\tau_i =-T}^{\tau_f=T} X(\tau)d\tau = 0$ if we require the
fluctuation to be orthogonal to
the zero mode $\frac {d\phi_c}{d\tau}$, i.e.$\int_{-T}^{T} \frac
{d\phi_c(\tau)}{d\tau} X(\tau) d\tau = 0$, so that
$X$ has to satisfy $X(\tau) = -X(-\tau)$.
The above equations  therefore imply that $C_4=C_5$.  The previous equation
therefore implies that
$C_4 =C_5=0$.
 Using the shift--method--transformation\cite{Dittrich} we can set
\begin{equation}
X(\tau)= Y(\tau) + N(\tau)\int_{\tau_i}^\tau \frac {\dot{N}(\tau')}{N^2(\tau')}
Y(\tau') d\tau'
\end{equation}
with
\begin{equation}
N(\tau)\equiv \frac{d\phi_c}{d\tau} = \frac {kb^2(k)a}{\mu}
cn[b(k)\tau]dn[b(k)\tau]
\label{equation(8.4)}
\end{equation}
and
\begin{equation}
I= \frac{1}{\sqrt{2\pi}} [\frac{1}{N(\tau_i)N(\tau_f)\int _{\tau_i}^{\tau_f}
\frac{d\tau}{N^2(\tau)}}]^
{\frac{1}{2}}
\label{26}
\end{equation}
This expression is singular at the turning point values of $\tau_f$ and
$\tau_i$,                                     since the
``velocities''expressed by the zero modes vanish at the
turning points;this is different from the case of vacuum instantons or vacuum
bounces in
which case the turning points can be reached only asymptotically.  Our
procedure here is
to use the end-point integrations in the expression for the transition
amplitude in order
to smooth out the singularities in $I$. One can show that in approaching the
limits
$\tau_{f,i} \rightarrow \pm T$ the following expression holds
formally\cite{Liangthree}
\begin{equation}
\frac{\partial^2 S_c}{\partial \phi^2(\tau_f)} = [N(\tau_f)N(\tau_i)
\int_{\tau_i}^{\tau_f}\frac{d\tau}{N^2(\tau)}]^{-1}
\end{equation}
In the integral
\begin{equation}
A_{+,-}=\int
\psi_{E+}(\phi_f)\psi_{E-}(\phi_i)\exp{[-S_c(\phi_f,\phi_i,\tau_f-\tau_i)]}
I(\phi_f,\phi_i,\tau_f,\tau_i)d\phi_f d\phi_i
\end{equation}
we replace the wave functionals by their respective WKB approximations  in the
barrier
(from $\phi(-T)=-\tilde{a} $ to $ \phi(T) = \tilde{a} $)
, i.e.we set
\begin{eqnarray}
\psi_{E+}(\phi_f) &=&\frac{C_+\exp{[-\Omega (\phi_f)]}}{\sqrt{N(\tau_f)}}
\equiv\frac{C_+
\exp{[-\int_{\phi_f}^{\tilde{a}}\dot{\phi}d\phi]}}{\sqrt{N(\tau_f)}}\nonumber\\
\psi_{E-}(\phi_i) &=& \frac{C_-\exp{[-\Omega
(\phi_i)]}}{\sqrt{N(\tau_i)}}\equiv\frac{C_-\exp{[-\int
 _{-\tilde{a}}^{\phi_i} \dot{\phi} d\phi]}}{\sqrt{N(\tau_i)}}\nonumber\\
\label{27}
\end{eqnarray}
where the constants $C_+, C_-$  can be calculated from integrals over the two
domains
$(-\tilde{a}', -\tilde{a})$ and $(\tilde{a},\tilde{a}')$ neighbouring the
barrier and  are given by
\begin{equation}
C_+=C_-=\left[{\frac{\frac{1}{2}}{\int_{\tilde{a}}^{\tilde{a'}}
\frac{d\phi}{\sqrt{2(E-V)}}}}\right]
^{\frac {1}{2}}
      =\left[{\frac{\mu \sqrt{1+u}}{2K(k')}}\right]^{\frac{1}{2}} \equiv C
\end{equation}
We are interested in the limits
\begin{equation}
\phi_i\rightarrow \phi(-T)\equiv -\tilde{a}, \phi_f \rightarrow \phi(T) \equiv
\tilde{a}'
\end{equation}
We therefore use Taylor expansion in $\phi_f$ around a nearby point
$\phi(\tau_0)$ so that
\begin{eqnarray}
\exp{[-S_c(\phi_f,\phi_i,\tau_f-\tau_i)]} & = & \exp{[-S_c(\phi(\tau_0),\phi_i,
\tau_0-\tau_i)]} \cdot
\nonumber\\ & & \cdot  \exp{[-\frac{1}{2}
(\frac {\partial^2S_c}{\partial {\phi_f}^2})_{\phi_f=\phi(\tau_0)}
(\phi_f-\phi(\tau_0))^2]}
\end{eqnarray}
and
\begin{equation}
\exp{[-\Omega(\phi_f)]}=\exp{[-\frac{1}{2} (\frac {\partial^2\Omega
(\phi_f)}{\partial \phi_f^2})_{\phi_f
=\phi(\tau_0)} (\phi_f-\phi(\tau_0))^2]}
\end{equation}
in the Gaussian approximation.
We also have with $\phi(\tau) = - \phi(-\tau)$
\begin{eqnarray}
(\frac {\partial^2S_c}{\partial \phi_f^2})_{\phi_f=\phi(\tau_0)} &=&- \frac
{\dot{N}(\phi(\tau_0))}{N(\phi(\tau_0))}
 +\frac {1}{N^2(\phi(\tau_0))\int_{-\tau_0}^{\tau_0} \frac
{d\tau}{N^2(\tau)}}\nonumber\\
(\frac {\partial^2\Omega}{\partial \phi_f^2})_{\phi_f=\phi(\tau_0)} &=& + \frac
{\dot{N}(\phi(\tau_0))}
{N(\phi(\tau_0))}\nonumber\\
\end{eqnarray}
(here the $\Omega$ contribution is seen to be one degree less divergent than
that of $S_c$)
and the relations
\begin{eqnarray}
S_c(\phi(T),\phi(-T),2T) &=& W(\phi(T), \phi(-T), E_c) + 2E_cT\nonumber\\
W &=& \frac{4\mu a^2}{3} (1+u)^{\frac {1}{2}} [E(k)-uK(k)]\nonumber\\
\end{eqnarray}
and write $\frac {d\phi_i}{N(\phi_i(\tau))}=d\tau$.The integration with respect
to $\phi_f$ becomes
Gaussian and can be carried out first. Then the limit $\tau_0 \rightarrow T$ is
taken.
Finally integrating with respect to $\tau $ from
$-T$ to $T$ we obtain for the amplitude in the one-loop approximation
\begin{equation}
A_{+,-}=2TC^2\exp{[-W]}\exp{[-2E_cT]} \equiv S_{+,-} \exp{[-2E_cT]}
\end{equation}
(For comparison with the S-matrix calculation to be mentioned below we note
here that
the factor $\exp[-2E_cT]$ represents the free field evolution part (here in
Euclidean time)
 so that in the limit of
vacuum boundary conditions  the remaining part $S_{+,-}$ can be looked at as
the
(here rather unconventional) S-matrix element in the one vacuum instanton
approximation
between the low-lying nth excited states in the two wells, i.e.$S_{+.-}\equiv
2TC^2 \exp [-W]$
for $k^2 \rightarrow 1)$.

Proceeding similarly in the case of amplitude contributions stemming from one
nonvacuum
instanton and  respectively one or n nonvacuum instanton pairs we obtain
\begin{eqnarray}
A_{+,-}^{(3)}&=&\int_{-T}^Td\tau_1\int _{-T}^{\tau_1}d\tau_2\int _{-T}^{\tau_2}
 d\tau C^3\exp{[-3W]}\exp{[-2E_cT]}\nonumber\\
&=&\frac {(2T)^3}{3!} C^3\exp{[-3W]}\exp{[-2E_cT]}\nonumber\\
A_{+,-}^{(2n+1)}&=& \frac{(2T)^{2n+1}}{(2n+1)!}
C^{2n+1}\exp{[-(2n+1)W]}\exp{[-2E_cT]}\nonumber\\
\end{eqnarray}
Summing over $n$ and comparing the expression with the expression for $A_{+,-}
$ at the
beginning, we obtain the WKB level-splitting
formula
\begin{equation}
\triangle E =\frac{\mu\sqrt{1+u}}{K(k')}\exp{[-W]}
\label{28}
\end{equation}

We define as weak coupling those values of $\mu a^2$ which are such
that $g^2\equiv\frac {1}{\mu a^2}<<1$.
In this limit the two minima of the potential are widely separated and the
central
barrier becomes very high.
 In the following we shall consider high energies as
those associated with high quantum states.  We therefore replace $E_c$ by the
oscillator
approximation $E_n=(n+\frac{1}{2})\omega$ where $\omega = 2\mu$ .  In that
approximation
we have $V(\phi)\simeq 2\mu^2(\phi-\phi_{\pm})^2, \phi_{\pm}= \pm a$
and $\int_{\tilde{a}}^{\tilde{a'}}\sqrt{2(E-V)}d\phi
=(n+\frac {1}{2})\pi$.  We consider separately the cases of low and high
energies.

$ Low\; energies $. We have $u=\sqrt{2E_c}/{a\mu} = 2g\sqrt{n+\frac{1}{2}},
 k^2=\frac {1-u}{1+u} =1-k'^2$.
The appropriate expansions (for small $u$ or $k'^2$) of the elliptic integrals
are
\begin{eqnarray}
E(k)&=&1+\frac {1}{2} k'^2\{\ln (\frac {4}{k'})- \frac {1}{2}\} +...\nonumber\\
K(k)&=&\ln (\frac {4}{k'}) + \frac {1}{4} k'^2\{\ln(\frac{4}{k'})-1\} +...
\end{eqnarray}
With these expansions we obtain
\begin{equation}
W=\frac {4}{3g^2}+2(n+\frac {1}{2})\ln(\frac {g}{4})+(n+\frac
{1}{2})\ln(n+\frac {1}{2})-(n+\frac {1}{2})
\end{equation}
and hence
\begin{equation}
\triangle E_n=
\frac{2\mu}{\pi}[\frac{2^4e}{g^2(n+\frac{1}{2})}]^{n+\frac{1}{2}}e^{-\frac
{4}{3g^2}}
\end{equation}
Using the Stirling relation
\begin{equation}
[\frac {e}{n+ \frac {1}{2}}]^{n+ \frac {1}{2}} \approx
\frac{\sqrt{2\pi}}{n!}\frac{e^{\frac {1}{2}}}
{(1+\frac{1}{2n})^{n+\frac{1}{2}}}\approx \frac{\sqrt{2\pi}}{n!}
\end{equation}
we see  that
\begin{equation}
\triangle E_n= \frac{2\sqrt{2}\mu}{n!\sqrt\pi}
[\frac{2^4}{g^2}]^{n+\frac{1}{2}}
e^{-\frac{4}{3g^2}}
\end{equation}
This expression agrees, as expected,with the WKB-equivalent
result\cite{Achuthan} in
the low energy limit, i.e. for $u=0$. We also observe that
under the condition $g^2(n+\frac{1}{2})<<1, k'\rightarrow 0$ the amplitude $A$
as well as
$\triangle E_n$ grow with energy (due to the second term in $ W $ above).

These low energy results agree with those of Bachas et al.\cite{Bachas} who
estimated the
S-matrix element $ S_{n \rightarrow n}$ for the vacuum instanton transition
from the
nth asymptotic oscillator state on one side of the barrier to the nth
asymptotic oscillator
state on the other side using the LSZ procedure. Thus, defining $\phi_{\pm}$ by
\begin{equation}
\phi_{\pm}:= \pm a - \phi_c(x) \rightarrow 0, \;\; x \rightarrow \pm \infty
\end{equation}
we can construct effective boson creation and annihilation Heisenberg operators
$\hat{a}_{\pm}^{\dagger},\hat{a}_{\pm}$ in the wells ``$+$" and ``$-$" given by
\begin{equation}
\hat{a}_{\pm} =-\frac{i}{\sqrt\mu} e^{2i\mu t}\stackrel{\leftrightarrow} {\frac
{\partial}{\partial t}}
\phi_{\pm}(x= it)\stackrel{\pm\infty}{ \longrightarrow } 4a \sqrt \mu
\end{equation}
These operators are such that in the asymptotic limits they correspond to the
harmonic
oscillator operators $\hat{a}$ in
\begin{equation}
\phi (t) = \frac{1}{\sqrt{2\omega}}[\hat{a} e^{-i\omega t} +\hat{a}^{\dagger}
e^{i\omega t}]
\end{equation}
with $\omega = 2\mu$. The transition amplitude through the central barrier
induced by a
vacuum instanton is then
\begin{eqnarray}
_+<1,out|1,in>_- &=& <0|\hat{a}_+\hat{a}_-^\dagger|0>\nonumber\\
&=& \lim_{t \rightarrow -\infty,t'\rightarrow+\infty}
(\frac{i}{\sqrt\mu} e^{2i\mu t'}\stackrel{\leftrightarrow} {\frac
{\partial}{\partial t'}})
(\frac{-i}{\sqrt\mu} e^{-2i\mu t}\stackrel{\leftrightarrow}
{\frac{\partial}{\partial t}}) G\nonumber\\
&=&(4a\sqrt\mu)^2 I\nonumber\\
&=& S_{+,-}
\end{eqnarray}
where $ G= <0|\phi_+(x')\phi_-(x)|0> $ and $ I $ is the vacuum instanton
tunneling propagator

$ High\; energies $. High energies here means those approaching the top of the
barrier,the latter
generally being called the sphaleron mass.  In the present context this implies
$k\rightarrow 0$ and $E\rightarrow \frac{ a^2\mu^2}{2}$.  In this limit
\begin{equation}
W\approx \frac{\sqrt{2}\pi}{g^2}k^2 \rightarrow 0
\end{equation}
and
\begin{equation}
\frac{\mu\sqrt{1+u}}{K(k')}\approx
\frac{\sqrt{2}\mu}{\ln(\frac{4}{k})}\rightarrow 0
\end{equation}
Thus at these energies the amplitude and the splitting $\triangle E$ are no
longer
suppressed by the typical vacuum instanton factor $\exp(-\frac{4}{3g^2})$ but
by the
prefactor $\frac{\mu\sqrt{1+u}}{K(k')}\rightarrow 0$.Thus in the high energy
limit
\begin{equation}
A\sim\frac{\mu\sqrt{1+u}}{K(k')}\exp{[-W]}\rightarrow 0
\end{equation}
Similar results  can be expected for various other potentials, in particular
for the
sine-Gordon potential which has been considered in the
literature\cite{Liangfive}.

\section{Nonvacuum bounces and tunneling}
With methods similar to those described above one can consider an amplitude in
the
neighbourhood of a bounce  which is the classical configuration in the case of
the
inverted double--well potential \cite{Coleman,Bose,Liangtwo,Liangfour}
and calculate the imaginary part of the energy.  In this
case the small fluctuation equation has three negative eigenmodes, of which two
do not contribute to the amplitude in view of boundary conditions and the
remaining
one is responsible for the imaginary part.  For further details we refer
to the literature\cite{Liangtwo,Liangfour}.

\section{The Bogomolny-Fateyev relation and conclusions}
In the case of  systems with more than one classical ground state, the
classical vacuum
state chosen as the perturbation theory vacuum in general does not coincide
with the
true quantum mechanical ground state.  Thus although the exact ground state is
stable, the corresponding perturbation theory vacuum is only metastable due to
the
possibility of tunneling to the other vacuum states. For example, the
double-well
potential
\begin{equation}
V(\phi)=\frac{\lambda^2}{2}(\phi^2-\frac{1}{\lambda^2})^2
\end{equation}
leads to the level splitting calculated above.
But the following distorted form of the potential
\begin{eqnarray}
V(\phi)&=&\frac{1}{2}\lambda^2 (\phi^2 -\frac{1}{\lambda^2})^2 \;\;
\mbox{for}\; \;\phi\leq\frac{1}{\lambda}\nonumber\\
           &=&-\frac{1}{2}\lambda^2(\phi^2-\frac{1}{\lambda^2})^2
\;\;\mbox{for}\;\;
                  \phi>\frac{1}{\lambda}
\end{eqnarray}
results in an imaginary part of the energy, $ \Im m E$, for a ``real''
metastable ground state.
The shape of the potential is then similar to that of a cubic potential which
is
the easiest example of a potential with a bounce\cite{MK}.
(The unphysical shape of $V(\phi)$ for $\phi > \frac{1}{\lambda} $ is
irrelevant here;equivalently
one could assume a flat behaviour or even a rising one far away so that in any
case the tunneling
particle would behave as free over some distance sufficiently far away from
$\phi = \frac{1}{\lambda}$).
Bogomolny and Fateyev\cite{Bogomolny} observed that (to leading order)
\begin{equation}
\triangle E = 2 \pi i(\delta E)^2
\end{equation}
where $\triangle E$ is the discontinuity of the ground state energy at the cut
$\lambda^2 \geq 0$
with $\triangle E = 2 i \Im m E$ while $\delta E$ is the instanton contribution
to the real part
of the ground state energy (i.e. for the double-well potential), namely the
level shift due
to quantum tunneling.  The Bogomolny-Fateyev relation has been verified and
extended to excited
states by comparing the explicit expressions of the two quantities for both the
double-well
potential and the periodic potential and their appropriately distorted versions
in the above
sense \cite{Achuthan}. The formula serves as a crucial test of the validity of
calculating
quantum tunneling effects with nonvacuum instantons and nonvacuum
bounces.In the considerations above
the level splitting for the excited states of the double-well potential was
obtained with
nonvacuum instantons . The classical solution which extremises the Euclidean
action
is
\begin{equation}
\phi_c(\tau)= \frac{kb(k)}{\lambda} sn[b(k)\tau,k]
\end{equation}
where
\begin{equation}
k^2 =\frac{1-u}{1+u}, \; u=\lambda \sqrt{2E},\;  b(k) =
[\frac{2}{1+k^2}]^{\frac{1}{2}}
\end{equation}
The Jacobian elliptic function $sn(z,k)$ has period ${\cal T} = 4nK(k)$. In the
calculation of
the level splitting the solution for a half period is regarded as a nonvacuum
instanton configuration.
The level shift (i.e. half of the level splitting ) is obtained as
\begin{equation}
\delta E = B \exp [-W']
\end{equation}
where the prefactor $ B $  is given by
\begin{equation}
B = \frac{[1+u]^{\frac{1}{2}}}{2K(k')}
\end{equation}
and $ W' $ by
\begin{equation}
W' = \frac{4}{3\lambda^2} (1+u)^{\frac{1}{2}}[E(k) - uK(k)]
\end{equation}
If we regard the configuration over the full period  (effectively a nonvacuum
                                       instanton--anti-instanton pair) as a
bounce configuration which returns to its
original position (such a consideration has also been discussed by Hawking and
Ross
\cite{Hawking})we can write it
\begin{equation}
\tilde{\phi}_c(\tau) = \frac{k b(k)}{\lambda} sn [b(k)\tau + K(k),k]
\end{equation}
so that this is zero at $b(k)\tau = -K(k)$, i.e. at $\tau= -T$, and at
$b(k)\tau = K(k)$,
i.e. at $\tau = +T$ (since $sn u$ vanishes for $u = 0, 2K(k)$).
This motion is allowed for a physical system with the distorted potential
above. The motion
of the bounce starts at $\tau = - 2T$ and ends at $\tau = +2T$. Since $ sn[u +
K(k),k] =
cn[u,k]/dn[u.k] $ we see that this bounce is an even function of $u$.  The zero
mode $\psi_0$,
i.e. the derivative of the bounce (which corresponds classically to its
velocity), is therefore odd, i.e. a wave function
with eigenvalue zero which passes through zero at $ u=0$.  Thus the ground
state eigenfunction of
 the
corresponding fluctuation equation must have a negative eigenvalue. This
negative eigenvalue is the
one which is responsible
for the instability of the configuration. The imaginary
part of the energy is  now obtained the way we obtained it
elsewhere\cite{Liangtwo}.  We then have
\begin{equation}
\Im m E = B \exp [-2W']
\end{equation}
and so
\begin{equation}
\Im m E = \frac{1}{B} (\delta E)^2
\end{equation}
In the low energy limit $\frac{1}{B} = \pi $, and the Bogomolny--Fateyev
relation holds exactly.
We conclude, therefore, that the consideration of classical finite energy
nonvacuum configurations
is applicable to numerous tunneling phenomena. Given $\delta E$ one can use the
Bogomolny--
Fateyev relation (by inserting $\Im m E $ into the appropriate moment integral)
in order to
derive the behaviour of a large order term of the perturbation expansion (in
Borel nonsummable
cases) of the eigenvalue $ E $ in the case with splitting.  The
Bogomolny--Fateyev relation
has also been observed in other related contexts\cite{Zinn} and
computationally\cite{Damburg}.
The contribution of sphalerons to the large-order behaviour of perturbation
expansions
in quantum mechanical models derived from nonlinear sigma models with
symmetry--breaking
potentials has been investigated recently by Rubakov and Shvedov\cite{Rubakov}.
We also mention that it is possible to develop a BRST-invariant approach to
quantum
mechanical tunneling which avoids the degeneracy problem of ill defined path
integrals due to
zero modes. So far we have applied this method only to the sine--Gordon
potential\cite{Zhou}.
Finally we remark that sphaleron configurations analogous to those discussed
here
arise also in other theories such as Skyrme-like models\cite{Tchrakian} and
Yang-Mills
and sigma-model theories\cite{Obrien}.

\end{document}